\UseRawInputEncoding
\documentclass[12pt]{article}
\usepackage{}
\usepackage{geometry}             
\usepackage{graphicx}
\usepackage{amssymb}
\usepackage{amsmath}
\usepackage{epstopdf}
\usepackage{comment}
\usepackage{cite}
\usepackage{caption}
\usepackage{subfigure}
\usepackage{epstopdf}

\usepackage[hyperindex=true,
          pdfstartview=FitH,
          bookmarksnumbered=true,
          bookmarksopen=true,
          citecolor=blue,
          linkcolor=blue,
          colorlinks=true,
          pdfborder=001,
          unicode]{hyperref}

\allowdisplaybreaks

\begin{document}

\title{Hawking-Page transition with reentrance and triple point in Gauss-Bonnet gravity}
\author{Yuan-zhang Cui$^{1}$\thanks{{\em
        email}: \href{mailto:kylins@cug.edu.cn}{kylins@cug.edu.cn}}, Wei Xu$^{1}$ \thanks{{\em
        email}: \href{mailto:xuwei@cug.edu.cn}
        {xuwei@cug.edu.cn}} \thanks{Corresponding author}, and Bin Zhu$^{2,3}$\thanks{{\em
        email}: \href{mailto:zhubin@mail.nankai.edu.cn}
        {zhubin@mail.nankai.edu.cn}}\\
$^{1}$School of Mathematics and Physics,\\
China University of Geosciences, Wuhan 430074, China\\
$^{2}$Department of Physics, Yantai University, Yantai 264005, China\\
$^{3}$Department of Physics, Chung-Ang University, Seoul 06974, Korea}
\date{}
\maketitle
\begin{abstract}
In this paper, a new family of Hawking-Page (HP) transition, the HP transition with reentrance and triple point
is introduced for the first time, by investigating HP transition of hyperbolic AdS black hole in extended
thermodynamics of general dimensional Gauss-Bonnet gravity.
The Reentrant HP transition is composed of two HP transitions with a large and a small HP temperature,
and the triple point corresponds to small black holes/massless black holes/large black holes (SBHs/MBHs/LBHs) phases all coexisting.
We present the two branches of HP transition temperatures,
which both depend on the pressure (i.e. the cosmological constant) and the Gauss-Bonnet constant.
Besides, the pressure and the Gauss-Bonnet constant both enlarge the large
HP temperature and diminish the small HP temperature.
We also show the coexistence lines in the $P-T$ phase diagrams.
The triple point and an up critical point in phase diagrams for arbitrary dimensional Gauss-Bonnet AdS black hole systems are given, together with some interesting universal relations which only depend on the dimensions of spacetime.
We can explain the reentrant HP transition and triple point of SBHs/MBHs/LBHs
as the triple point of Hadronic matter/Quark-Gluon/Plasma/Quarkyonic matter in the QCD phase diagram.
These results may improve the comprehension of the black hole thermodynamics in the quantum gravity framework and shed some light on the AdS/CFT correspondence beyond the classical gravity limit.
\end{abstract}

\section{Introduction}
Black hole thermodynamics has attracted a great attentions in gravitational theory,
as it is widely believed that black hole thermodynamics could provide a further insight into the understanding of quantum gravity.
Especially, the famous Hawking-Page (HP) transition \cite{Hawking:1982dh} starts an exciting period of exploring the holographic and quantum understanding of critical phenomena and phase transition in general AdS spacetime.
Afterwards, R. Emparan, et al, introduce a first order phase transition of RN-AdS black hole \cite{Chamblin:1999tk,Chamblin:1999hg}
similar to the liquid/vapour phase transition of the van der Waals fluid.
After treating the cosmological constant as a thermodynamic pressure
\cite{Kastor:2009wy,Dolan:2011xt,Cvetic:2010jb,Dolan:2013ft,Kastor:2010gq,Castro:2013pqa,El-Menoufi:2013pza},
the small black holes/large black holes phase transition of RN-AdS black hole is established \cite{Kubiznak:2012wp},
which is precisely analogous to the liquid/vapour phase transition of the van der Waals fluid
(See also \cite{Altamirano:2014tva,Kubiznak:2016qmn} for reviews).
Other interesting families of phase transition are also studied in black hole thermodynamics,
including the Reentrant phase transition \cite{Gunasekaran:2012dq,Altamirano:2013ane} and superfluid phase transition \cite{Hennigar:2016xwd}.

The HP transition characterizes a first order phase transition occurs between a large AdS black hole
and a thermal AdS vacuum \cite{Hawking:1982dh}.
Since the AdS black hole phase dominates the partition function at a high temperature limit,
while the thermal AdS vacuum dominates at a low temperature limit,
the thermal AdS gas will collapse to a stable large black hole when the temperature increases.
Especially, The HP transition
could be explained as the confinement/deconfinement phase transition of gauge field \cite{Witten:1998zw}, inspired by the
AdS/CFT correspondence \cite{Maldacena:1997re,Gubser:1998bc,Witten:1998qj}.
Very recently, many studies about HP transition in different backgrounds are aroused \cite{Mbarek:2018bau,Xu:2019xif,Aharony:2019vgs,Wu:2020tmz,Wang:2019vgz,Lala:2020lge,Wang:2020pmb,Zhao:2020nrx,Yan:2021uzw,Su:2021jto,Du:2021wvt}. Especially, there are some viewpoints about the HP transition in the microcosmic  \cite{Li:2020khm,Xu:2020ubo,Li:2021zep} and holographic framework \cite{Copetti:2020dil,Wei:2020kra,Belhaj:2020mdr}.
Thus it is really important to investigate the HP transition,
since it should improve our understanding of the quantum and holographic properties of gravity in the spirit of the
AdS/CFT correspondence.

The studies about the HP transition also induce many
applications about the mutual study on the particle physics, especially in the QCD phase diagram.
Beyond the confinement/deconfinement phase transition,
the crossover is also speculated that, in the holographic dictionary,
corresponds to the HP crossover of Schwarzschild AdS black hole \cite{Nicolini:2011dp} in noncommutative spacetime
which is thought to be an effective description of quantum gravitational spacetime.
This crossover is the critical endpoint of the deconfinement phase transition \cite{Fromm:2011qi}, which is a continuous phase transition.
On the other hand, after considering the Quarkyonic matter \cite{McLerran:2007qj,Hidaka:2008yy,McLerran:2008ua,Fukushima:2008wg,Glozman:2007tv,DeTar:2009ef,Andronic:2009gj},
a phase transition composed of the deconfinement phase transition
and the quarkyonic transition, and a triple point, for Hadronic matter, Quarkyonic matter and the Quark-Gluon Plasma
are introduced. There are also some shreds of evidence from Lattice QCD simulation \cite{DeTar:2009ef}.
Many recent studies present the existence of this phase transition and triple point \cite{Andronic:2009gj,McLerran:2007qj,Doi:2014zea,Pak:2015dxa,Suganuma:2017syi,Yang:2020hun,McInnes:2009zp,McInnes:2009ux,McInnes:2010ti,McInnes:2012bt,Ong:2014maa,McInnes:2015pya,McInnes:2015hga,McInnes:2016dwk,McInnes:2018mwj,Henriksson:2019ifu}.
It is natural to ask,
how can the this phase transition and triple point be explained in the gravity side, in the AdS/CFT correspondence?

The answer should be the reentrant phase transition, which is composed of at least two phase transition.
The reentrant phase transition has previously been observed in a nicotine/water mixture \cite{Hudson1904}, granular superconductors, liquid crystals, binary gases, ferroelectrics and gels (see \cite{Narayanan1994} and the references therein).
In black hole thermodynamics, the reentrant phase transition is also presented in \cite{Gunasekaran:2012dq,Altamirano:2013ane}, which always appeared simultaneously with the triple point in the phase diagrams.
Therefore, we focus on a new family of HP transition, the HP transition with reentrance and triple point in Gauss-Bonnet gravity.
This reentrant HP transition is found for the first time, by investigating HP transition of hyperbolic AdS black hole in extended thermodynamics of $d$ dimensional Gauss-Bonnet gravity, which is composed of two HP transitions.
Many typical features of the reentrant HP transition are given.
We will discuss the duality between the reentrant HP transition and the triple point for small black holes/massless black holes/large black holes (SBHs/MBHs/LBH) and the phase transition and the triple point for Hadronic
matter/Quark-Gluon Plasma/Quarkyonic matter in QCD phase diagram.
These results may improve our understanding of the black hole thermodynamics in the quantum gravity framework and shed some light on the AdS/CFT correspondence beyond the classical gravity limit.

The paper is organized as follows: we revisit the extended thermodynamics of hyperbolic AdS black hole in $d$ dimensional Gauss-Bonnet gravity in the next section. In section 3 and 4, we study the HP transition with reentrance and triple point
in four and $d\geq5$ dimensions, respectively.
In section 5, we discuss the possible relevance and interpretation of the reentrant HP transition in the CFT picture. Finally some concluding remarks are given.

\section{Extended thermodynamics of hyperbolic AdS black hole in Gauss-Bonnet gravity}
In this section, we revisit the extended thermodynamics of hyperbolic AdS black hole in $d$ dimensional Gauss-Bonnet gravity.
This black hole solution is well known to take the form \cite{Boulware:1985wk,Wiltshire:1985us,Cai:2001dz,Cvetic:2001bk}
\begin{align}
    \mathrm{d} s^2&=-f(r)\mathrm{d} t^2+\frac{1}{f(r)}
  \mathrm{d} r^2+r^2\mathrm{d} \Omega_{d-2,k}^2,\\
  f(r)&=\frac{r^2}{2 \alpha } \left(1-\sqrt{\frac{64 \pi  \alpha  M }{(d-2)r^{d-1}}-\frac{64 \pi  \alpha  P}{(d-2) (d-1)}+1}\right)+k,
\end{align}
where $\mathrm{d} \Omega_{d-2,k}^2$ is the line element of a $(d-2)$-dimensional maximally symmetric Einstein manifold with curvature $k=-1$ corresponding to the hyperbolic topology of the black hole horizon.
$M$ is the mass of the black hole. $\alpha=(d-3)(d-4)\alpha_{GB}$ is a renormalized cosmological Gauss-Bonnet coupling constant, while $\alpha_{GB}$ is the real Gauss-Bonnet coupling constant.
In this paper, we will consider only the case $\alpha>0$, i.e. $\alpha_{GB}>0$,
since $\alpha_{GB}$ can be identified with the inverse string tension with positive value
if the theory is incorporated in string theory \cite{Boulware:1985wk}.
We will take the spacetime dimension $d\geq4$, since in $d=4$ dimensions,
though the Gauss-Bonnet term is a topological invariant and does not contribute to the spacetime,
it has a notable effect on black hole thermodynamics resulted from a non-trivial black hole entropy.
Finally, in the black hole chemistry framework, $P=-\frac{\Lambda}{8 \pi}=\frac{(d-1)(d-2)}{16\pi\ell^2}$  is the thermodynamic pressure associated with the cosmological constant $\Lambda$ \cite{Kastor:2009wy,Dolan:2011xt,Cvetic:2010jb,Dolan:2013ft,Kastor:2010gq,Castro:2013pqa,El-Menoufi:2013pza}, with $\ell$ being the $d$-dimensional AdS radius.

We list the thermodynamical quantities, including the black hole mass, temperature and entropy \cite{Cai:2013qga,Clunan:2004tb}
\begin{align}
			M&=\frac{r_{+}^{d-5} \bigg(\alpha  (d-1)(d-2) -(d-1)(d-2)  r_{+}^2+16 \pi  P r_{+}^4\bigg)}{16 \pi  (d-1)},\label{Td}\\
			T&=\frac{\alpha  (d-2)(d-5) -(d-2)(d-3) r_{+}^2+16 \pi  P r_{+}^4}{4 \pi  (d-2) r_{+} \left(r_{+}^2-2 \alpha \right)},\\
			S&=\frac{1}{4} r_{+}^{d-2} \left(1-\frac{2 \alpha  (d-2) }{(d-4) r_{+}^2}\right),
\end{align}
where $r_{+}$ is the event horizon radius of the black hole.
In black hole chemistry, to observe the HP transition, we present the Gibbs free energy
\begin{align}
  G&=H-TS=M-TS\nonumber\\
  &=\dfrac{-1}{16\pi(r_{+}^2-2\alpha)} \left( \frac{(d-2) \left(d^2-5 d-96 \pi  \alpha  P+4\right)r_{+}^{d-1} }{(d-4)(d-1)}\right.\nonumber\\
			&\left.-\frac{\alpha  (d-8) r_{+}^{d-3}}{(d-4)  (d-1)}+\frac{16 \pi  P r_{+}^{d+1}}{(d-2) (d-1)}+\frac{2 \alpha ^2 (d-2) r_{+}^{d-5}}{d-4}  \right),
\end{align}
which characterizes the canonical ensemble.
Here the black hole mass $M$ should be identified with the enthalpy $H$
rather than the internal energy of the gravitational system \cite{Kastor:2009wy}.

In what follows, we will focus on the famous HP transition,
for which a thermodynamically stable state is given by the global
minimum of $G$ of the black hole and the background spacetime (with zero Gibbs free energy).
To observe the phase transition, it is most useful to plot $G-T$ diagrams, fixing the other parameters.
This means that the case with a negative $G$ should be regarded as the Gauss-Bonnet black hole phase being
thermodynamically favored over the background spacetime;
and the case with vanishing $G$ just corresponds to the HP transition point,
which characterizes the phase transition between the Gauss-Bonnet black hole phase and the background
spacetime phase. Especially in this case, the background spacetime is the massless AdS black hole (MBH)
with a modified cosmological constant by the Gauss-Bonnet constant.
When the Gauss-Bonnet constant $\alpha$ is vanishing,
the spacetime reduces to the hyperbolic Schwarzschild AdS black hole,
which does not contain the HP transition.

\section{Reentrant HP transition and triple point in four dimensions}
In four dimensions, since the Gauss-Bonnet term is a topological invariant that does not contribute to the spacetime,
the spacetime becomes the hyperbolic Schwarzschild AdS black hole, i.e.
	\begin{equation}
			\begin{aligned}
				\mathrm{d}s^2&=-f(r)\mathrm{d}t^2+\frac{1}{f(r)}\mathrm{d}r^2+r^2\mathrm{d}\Omega^2_{2,-1},\\
				f(r)&=-\frac{2 M}{r}+\frac{8}{3} \pi  P r^2-1,
			\end{aligned}
	\end{equation}
which makes the discussion about the HP transition more clear than the cases in higher dimensions.
The thermodynamical quantities, mass and temperature of the system reduce to	
\begin{align}
	M=&\frac{4\pi r_{+}^3}{3} P -\frac{1}{2} r_{+}\\
	T=&2 P r_{+}-\frac{1}{4 \pi  r_{+}},\label{tem4}
\end{align}
which is exactly the same with the Schwarzschild AdS black hole;
while the black hole entropy breaks the area law and takes the form \cite{Castro:2013pqa,Fan:2014ala}
\begin{align}
  S=&\pi  \left(r_{+}^2-4 \alpha \right).
\end{align}
Thanks to this non-trivial black hole entropy,
the Gauss-Bonnet term has a notable effect on black hole thermodynamics even in four dimensions.
For example, the Gibbs free energy becomes
\begin{align}
  G=&-\frac{2}{3} \pi  P r_{+}^3+r_{+} \left(8 \pi  \alpha  P-\frac{1}{4}\right)-\frac{\alpha }{r_{+}}.
\end{align}

\begin{figure}[h!]
\begin{center}
\includegraphics[width=0.3\textwidth]{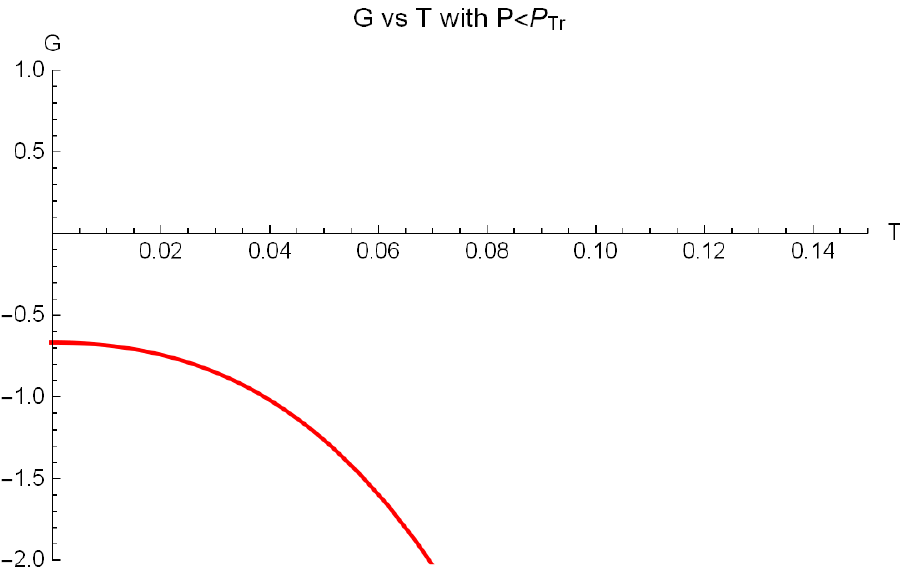}
\includegraphics[width=0.3\textwidth]{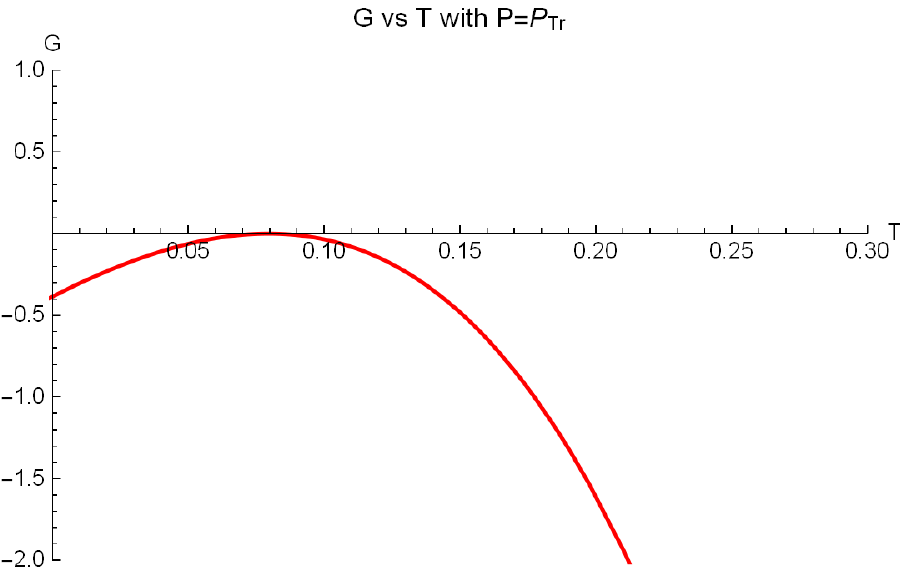}
\includegraphics[width=0.3\textwidth]{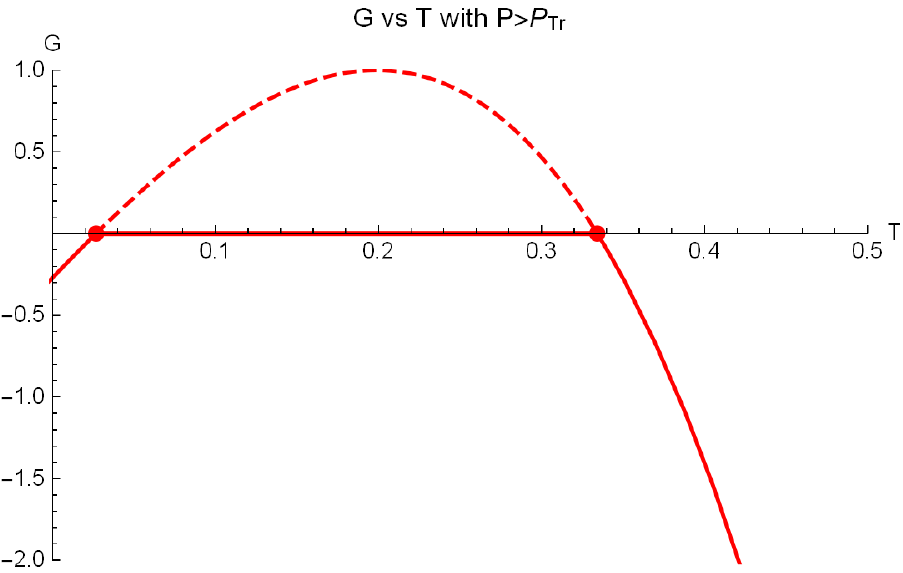}
\caption{Gibbs free energy vs temperature with different pressures (and $\alpha=1$) in four dimensions.
The globally stable states of the system are denoted by the thick lines.
When $P>P_{Tr}$, Black holes undergo the reentrant HP transition.}
\label{fig1}
\end{center}
\end{figure}

To consider the HP transition, we explore the global stability of Gibbs free energy.
This could be seen in the $G-T$ diagrams as shown in Fig.\ref{fig1}.
One can find that the existence of the HP transition depends on the maximum of the Gibbs free energy.
There is a critical pressure $P_{Tr}$ (We denote it as $P_{Tr}$ since it just corresponds to the triple point in phase diagram as shown later.), for black hole with $P=P_{Tr}$, the maximum of the Gibbs free energy becomes zero,
hence $G\leq0$; while for black holes with $P<P_{Tr}$, the Gibbs free energy is always negative.
This indicates that, when $P\leq P_{Tr}$, the hyperbolic AdS black hole phase is
therefore globally preferred than the background spacetime phase, and the HP transition will not happen here.
For black holes with $P>P_{Tr}$, one can see the right one of Fig.\ref{fig1}.
The maximum of the Gibbs free energy of black holes always divides the black holes into two branches.
It is easy to check that the temperature is a monotonically increasing function of mass,
thus we can denote the two branches of black holes as:
the small black holes (SBHs) with smaller mass and the large black holes (LBHs) with larger mass.
It is obvious that the situation becomes interesting, since there exist two zero free energy points.
The globally stable states of the system are denoted by the thick lines.
It is shown that here the two zero free energy points both correspond to the HP transitions,
with a large HP temperature and a small HP temperature.
If the temperature is smaller than the small HP temperature,
the background spacetime, i.e. the MBHs, should collapse into the SBHs;
while the MBHs should collapse into the LBHs when the temperature is larger than the large HP temperature;
because for these cases, the free energy of the black holes is lower than that of the background spacetime.
When the temperature stays in the intermediate region between the large and small HP temperature,
the system favors the MBHs phase.
Therefore, Black holes undergo the SBHs-MBHs-LBHs HP transition in this region of pressures.
This behavior is known as RPT, which is composed of two first order phase transitions (HP transition) in our case.

\begin{figure}[h!]
\begin{center}
\includegraphics[width=0.5\textwidth]{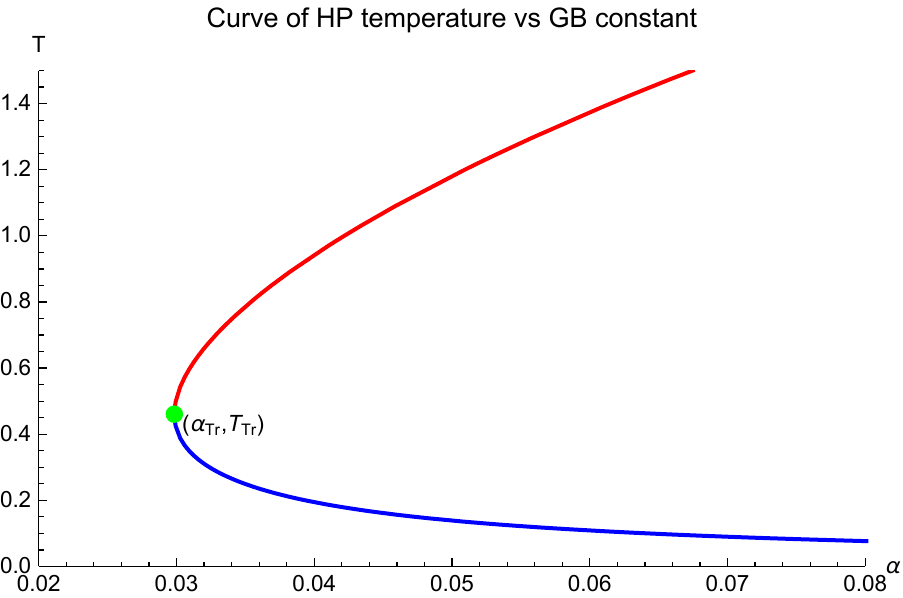}
\caption{The two branches of HP temperature vs Gauss-Bonnet constant with $P=1$ in four dimensions. The Gauss-Bonnet constant $\alpha$ enlarges the large HP temperature and diminishes the small HP temperature.}
\label{fig1-1}
\end{center}
\end{figure}

Now we calculate the HP temperature. After choosing a zero Gibbs free energy,
we can get the black hole radius of the HP transition
 	\begin{equation}
		r _{HP}=\frac{1}{4\sqrt{\pi P}} \sqrt{96 \pi  \alpha  P \pm \sqrt{(3-96 \pi  \alpha  P)^2-384 \pi  \alpha  P}-3}.\label{rHP}
	\end{equation}
From Eq.(\ref{tem4}), we can obtain the HP temperature
\begin{align}
  T_{HP}=T|_{r_+=r_{HP}}.
\end{align}
Since $r _{HP}$ should be positive, we find that the (reentrant) HP transition requires for an additional condition
	\begin{equation}
		96 \pi  \alpha  P \pm \sqrt{(3-96 \pi  \alpha  P)^2-384 \pi  \alpha  P}-3\geq 0,
	\end{equation}
which could be simplified as
\begin{align}
  \alpha P \geq \dfrac{3}{32\pi }.
\end{align}
The general behavior of the two branches of HP temperature are illustrated in Fig.\ref{fig1-1}.
It is clear that the (reentrant) HP transition only arises when $\alpha \geq \dfrac{3}{32\pi P}$.
Especially for the Gauss-Bonnet AdS black holes with $\alpha <\dfrac{3}{32\pi P}$ ,
the HP transition does not occur, which is consistent with the discussion about the hyperbolic AdS black hole in Einstein gravity (i.e. $\alpha=0$).
Besides, the Gauss-Bonnet constant $\alpha$ enlarges the large HP temperature and diminishes the small HP temperature.

\begin{figure}[h!]
\begin{center}
\includegraphics[width=0.5\textwidth]{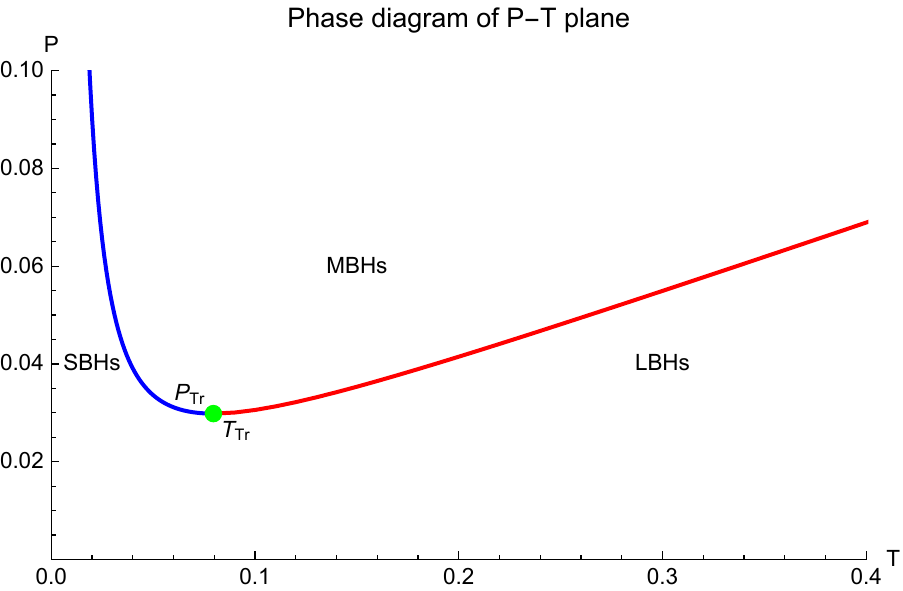}
\caption{The coexistence line: $P-T$ phase diagram (with $\alpha=1$) in four dimensions. The triple point is highlighted.}
\label{fig1-2}
\end{center}
\end{figure}

We finally show the coexistence line to have a whole picture of the reentrant HP transition.
This is plotted in $P-T$ phase diagram as shown in Fig.(\ref{fig1-2}).
When the temperature of the system is fixed, there always exists a single HP transition:
the SBHs/MBHs HP transition at low temperature or the LBHs/MBHs HP transition at high temperature.
When the pressure of the system is fixed, the reentrant HP transition arises,
since the system undergoes two HP transitions:
the SBHs/MBHs HP transition at low temperature and the LBHs/MBHs HP transition at high temperature.
The regimes observed in the $P-T$ phase diagram and described above could be understood as arising from a triple point
where SBHs, MBHs, and LBHs all coexist.
This triple point is located where the pressure of HP transition is reaching its limiting value.
Hence it is easy to derive the triple point by making the two $r_{HP}$ in Eq.(\ref{rHP}) coincide,
i.e.
	\begin{equation}
		(3-96 \pi  \alpha  P)^2-384 \pi  \alpha  P= 0.
	\end{equation}
We get the pressure and temperature of the triple point
\begin{align}
   P_{Tr}=\dfrac{3}{32\pi \alpha},\quad
   T_{Tr}=T_{HP}|_{P=P_{Tr}}=T|_{r_+=r_{Tr},P=P_{Tr}}=\dfrac{1}{4\pi \sqrt{\alpha}},
\end{align}
with the black hole radius $r_{Tr}=r_{HP}|_{P=P_{Tr}}=2\sqrt{\alpha}$;
while another one has a negative thus un-physical black hole radius.
Noting that only when $P> P_{Tr}$, there is a reentrant HP transition,
while the HP transition is vanishing when $P\leq P_{Tr}$;
which is consistent with the discussion in the $G-T$ diagrams.
Namely, the triple point exactly corresponds to the black hole phase whose maximum of the Gibbs free energy is zero.
Moreover, it is interesting to introduce a universal relationship of the triple point
\begin{align}
  \frac{P_{Tr}r_{Tr}}{T_{Tr}}=\frac{3}{4}.
\end{align}
	
\section{Reentrant Hawking-Page transition and triple point in $d\geq5$ dimensions}
The Hawking-Page transition in higher dimensions exhibits some differences from the case in four dimensions.
One can only see the Reentrant HP transition behavior of Gauss-Bonnet AdS black holes for
a range of pressure $P\in(P_{Tr},P_c)$.
In this section, we will firstly present the Reentrant Hawking-Page transition in five and six dimensions;
in other dimensions $d\geq5$ the behavior is similar.
Then we will derive the triple point and some universal relations from HP transition in $n$ dimensions.

\subsection{Reentrant Hawking-Page transition in five and six dimensions}
\begin{figure}[h!]
\begin{center}
\includegraphics[width=0.3\textwidth]{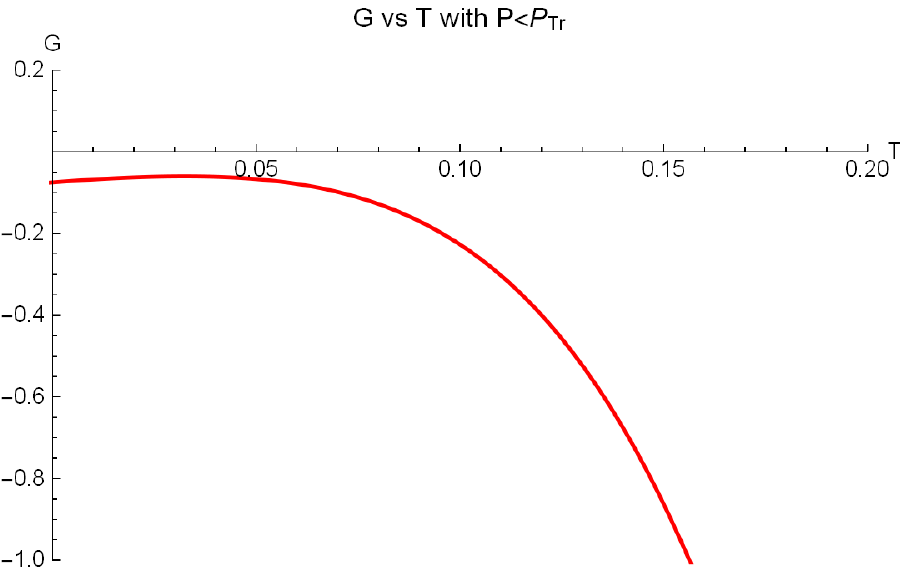}
\includegraphics[width=0.3\textwidth]{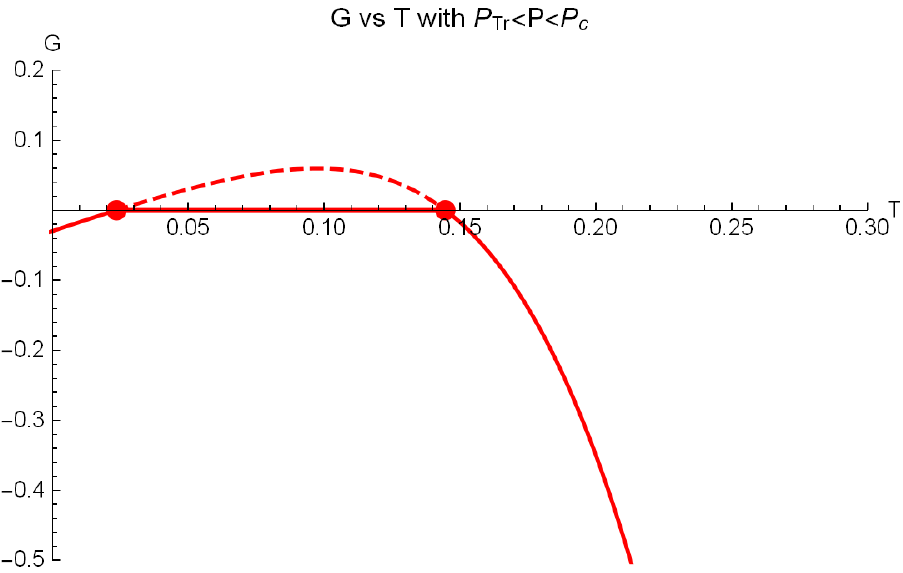}
\includegraphics[width=0.3\textwidth]{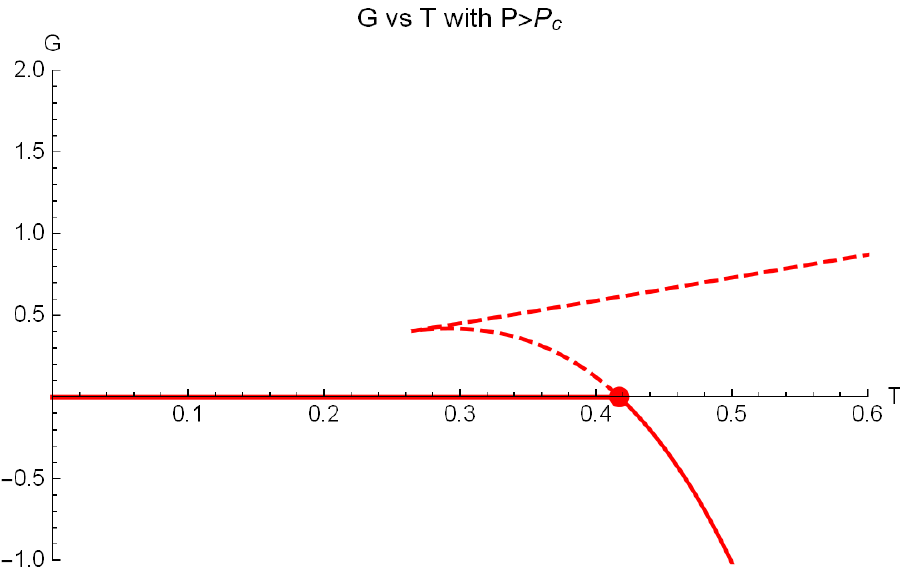}
\includegraphics[width=0.3\textwidth]{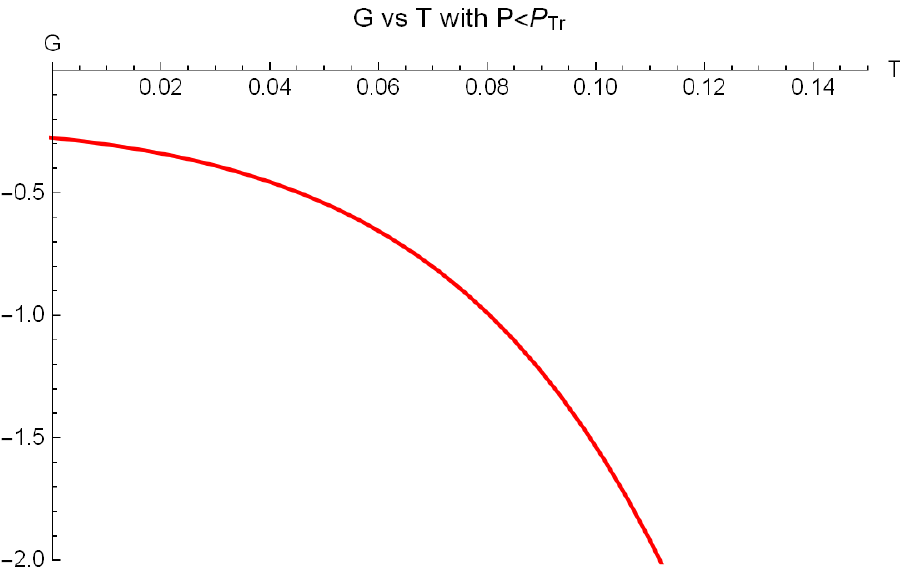}
\includegraphics[width=0.3\textwidth]{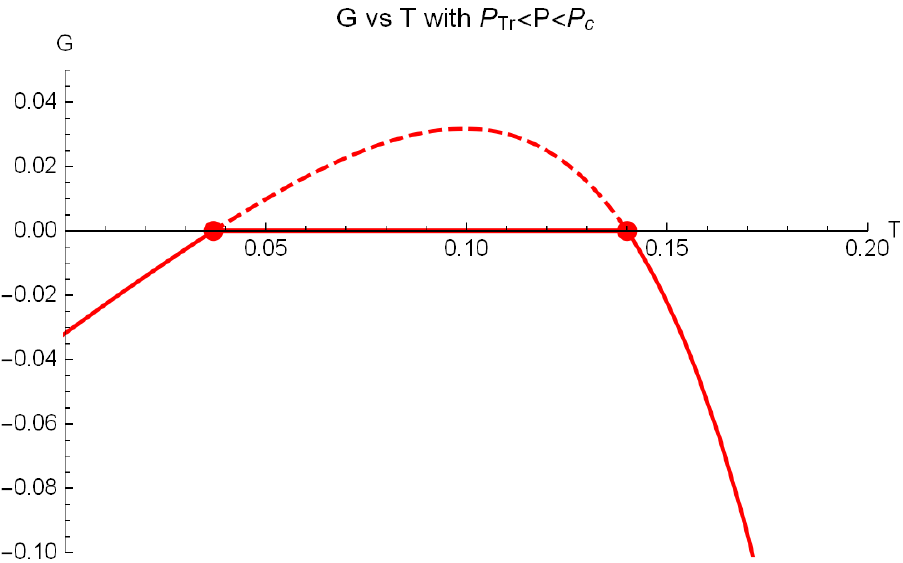}
\includegraphics[width=0.3\textwidth]{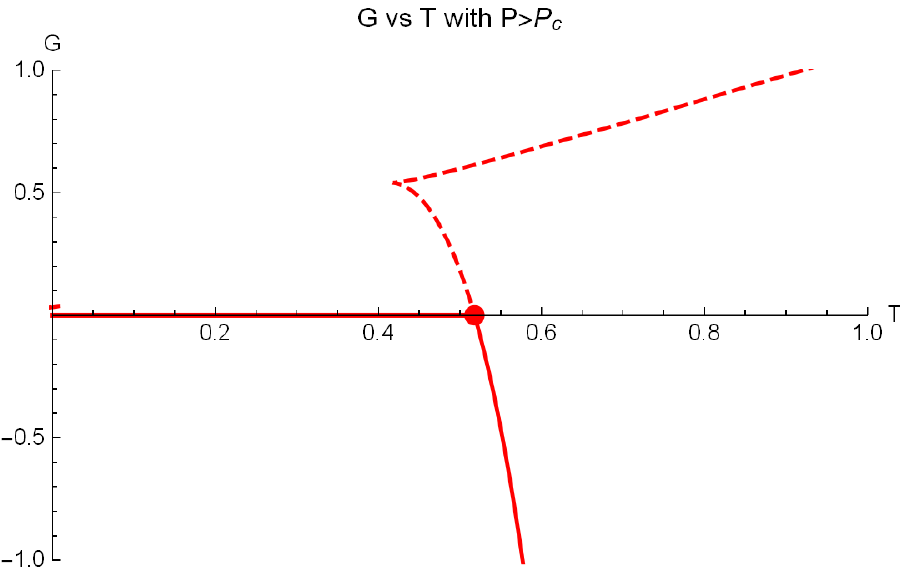}
\caption{Gibbs free energy vs temperature with different pressures (and $\alpha=1$) in five and six dimensions.
The above (below) three correspond to the cases in five (six) dimensions.
The globally stable states of the systems are denoted by the thick lines.
When $P_{Tr}<P<P_c$, Black holes undergo the reentrant HP transition; while there only exists a single HP transition when $P>P_c$.}
\label{fig2}
\end{center}
\end{figure}

In $d=5$ dimensions, the Gibbs free energy and temperature of Gauss-Bonnet AdS black holes reduce to
	\begin{equation}
		\begin{aligned}
			G=-&\frac{18 \alpha ^2+9 \alpha  r_{+}^2+(3-72 \pi  \alpha  P )r_{+}^4+4 \pi  P r_{+}^6}{48 \pi  (r_{+}^2-2 \alpha) }, \\
			T=&\frac{r_{+} \left(8 \pi  P r_{+}^2-3\right)}{6 \pi  \left(r_{+}^2-2 \alpha \right)}.
		\end{aligned}
	\end{equation}
In $d=6$ dimensions, the corresponding thermodynamical quantities become
	\begin{equation}
		\begin{aligned}
			G=&-\frac{r_{+} \left(20 \alpha ^2+5 \alpha  r_{+}^2+ (5-48 \pi  \alpha  P)r_{+}^4+4 \pi  P r_{+}^6\right)}{80 \pi  \left(r_{+}^2-2 \alpha \right)},\\
			T=&\frac{\alpha-3 r_{+}^2 +4 \pi  P r_{+}^4}{4 \pi r_{+} ( r_{+}^2-2 \alpha)}.
		\end{aligned}
	\end{equation}
The general behavior of the Gibbs free energy is illustrated in Fig.(\ref{fig2}), which have been plotted for
different pressures in $G-T$ diagrams. The globally stable states of the systems are denoted by the thick lines.
When the pressure is small ($P\leq P_{Tr}$), the Gauss-Bonnet AdS black hole phase always has a smaller Gibbs free energy than the background spacetime phase, which indicates no HP transition.
When the pressure becomes big ($P_{Tr}<P< P_{c}$), the reentrant HP transition, composed of two HP transitions, emerges,
as shown in the middle diagram of Fig.(\ref{fig2}).
From the right diagram of Fig.(\ref{fig2}), one can easily observe that
there is a single HP transition when the pressure becomes bigger ($P> P_{c}$),
since the temperature of another HP transition diverges or becomes negative.
(Noting that these ranges will be discussed later.)

\begin{figure}[h!]
\begin{center}
\includegraphics[width=0.4\textwidth]{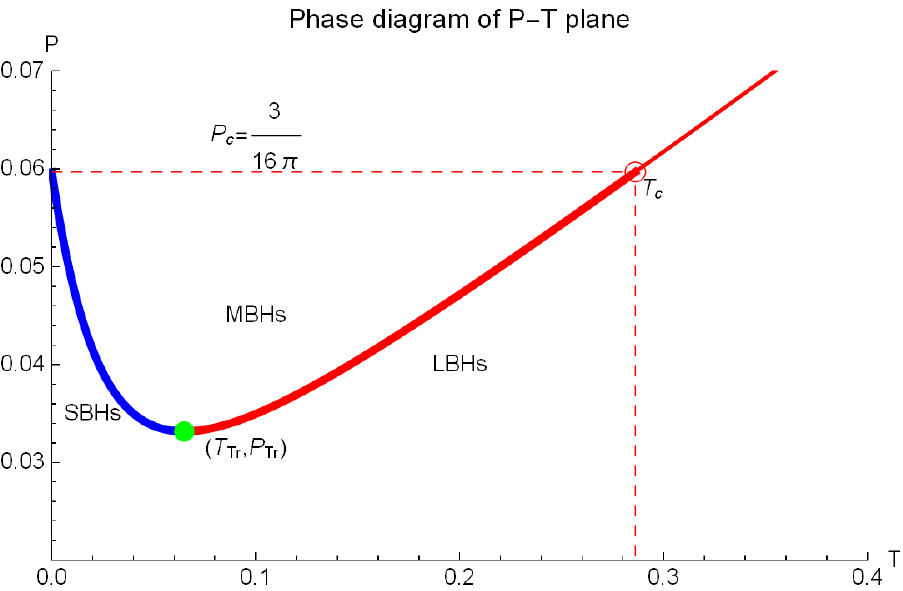}
\includegraphics[width=0.4\textwidth]{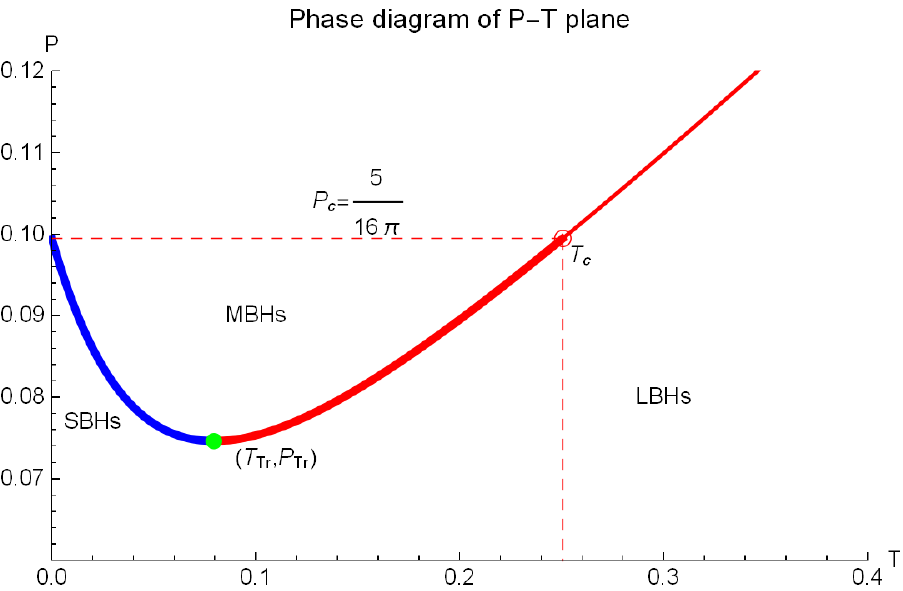}
\caption{The coexistence line: $P-T$ phase diagrams (with $\alpha=1$) in five and six dimensions
are plotted in the left and right figure, respectively.
The reentrant HP transition are denoted by the thick lines.
The triple points and up critical points are highlighted.}
\label{fig3-1}
\end{center}
\end{figure}

When we see the coexistence lines in the $P-T$ phase diagrams for the HP transition in five and six dimensions in Fig.(\ref{fig3-1}), the phase structure become more clear.
There is two branches of HP transition.
When a black hole crosses the solid line from left to right or bottom to top in the left branch,
it undergoes a HP transition from SBHs to MBHs; while when a black hole crosses the solid line from right to left or bottom to top in the right branch, it undergoes a HP transition from LBHs to MBHs.
The reentrant HP transition are denoted by the thick lines,
which could be bounded by the pressure of a triple point and an up critical point.
Namely, the reentrant HP transition behavior of Gauss-Bonnet AdS black holes
only exists for a range of pressure $P\in(P_{Tr},P_c)$.
Especially, $(P_{Tr},T_{Tr})$ denote the pressure and temperature of the triple point,
where SBHs, MBHs, and LBHs all coexist.
Therefore, we can derive the triple point by the property that
the triple point exactly corresponds to the black hole phase whose maximum of the Gibbs free energy is zero,
which will be shown in the next subsection.
Moreover, from Fig.(\ref{fig3-1}), one can also see that for Gauss-Bonnet AdS black holes with fixed Gauss-Bonnet constant $\alpha$, the pressure $P$ enlarges the large HP temperature and diminishes the small HP temperature.

We denote the pressure and temperature of the up critical point as $(P_{c},T_{c})$.
If the pressure approaches the critical value $P_c$, the left branch of the HP temperature will diverge,
other than become zero, which is evidently exhibited in the right diagram of Fig.(\ref{fig2}).
When the pressure is beyond the pressure of the up critical point,
the SBHs have a negative temperature which is physically unacceptable.
As a result, the left branch of the HP transition is vanishing,
and only the right branch is left.
Then one can never observe the reentrant HP transition.
We will derive the up critical point in the next subsection following the property that
the up critical point corresponds to the two-phase coexistence state having zero Gibbs free energy,
and its pressure $P_{c}$ coincides with the pressure of another two-phase coexistence state with a diverging HP temperature.

\begin{figure}[h!]
\begin{center}
\includegraphics[width=0.4\textwidth]{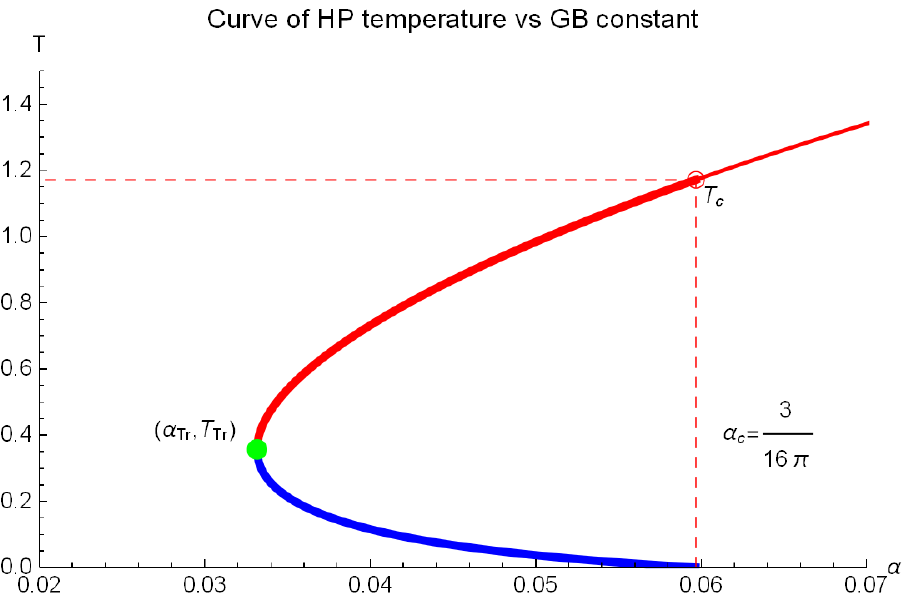}
\includegraphics[width=0.4\textwidth]{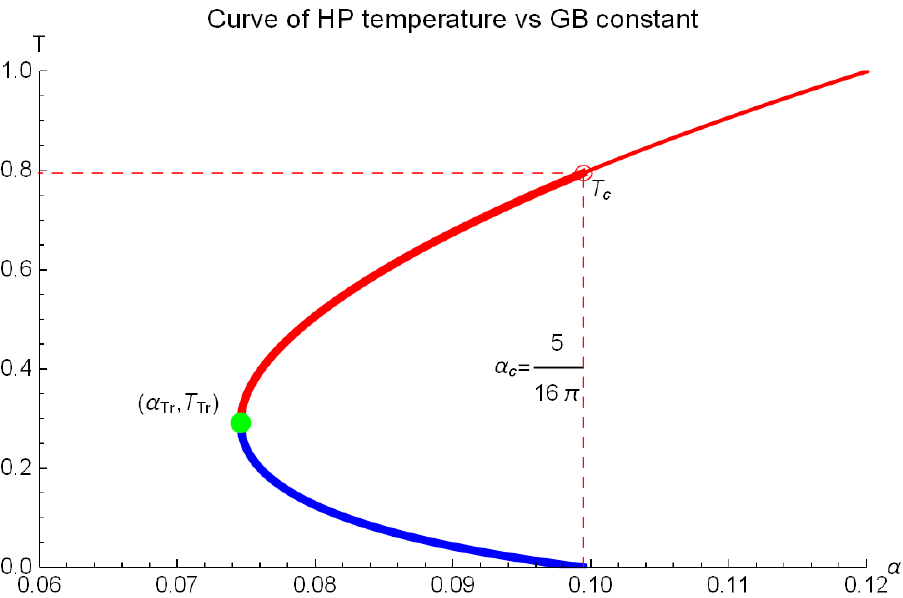}
\caption{The two branches of HP temperature vs Gauss-Bonnet constant with $P=1$ in five and six dimensions are shown in the left and right figure, respectively. The Gauss-Bonnet constant $\alpha$ enlarges the large HP temperature and diminishes the small HP temperature.}
\label{fig3}
\end{center}
\end{figure}

The two branches of HP temperature could be got from the zero Gibbs free energy as well,
which leads to
	\begin{equation}
		\label{f0}
		R^3+\frac{(d-2)  \left(d^2-5 d-96 \pi  \alpha  P+4\right)}{16 \pi  (d-4) P}R^2+\frac{\alpha  (d-8)(d-2)(d-1) }{16 \pi  (d-4) P}R+\frac{\alpha ^2 (d-2)^2 (d-1)}{8 \pi  (d-4) P}=0,
	\end{equation}
where $r_+=\sqrt{R}$ is inserted. This is a classical cubic equation whose roots can be analytically obtained.
Since the roots have a complicated form, we will not present them and the corresponding HP temperature here.
We also plot the two branches of HP temperature vs Gauss-Bonnet constant in five and six dimensions in Fig.(\ref{fig3}).
The HP temperature follows the similar property, as the Gauss-Bonnet constant $\alpha$ enlarges the large HP temperature and diminishes the small HP temperature. One should note that the HP transition with a big temperature only happens in Gauss-Bonnet AdS spacetime with $\alpha>\alpha_{Tr}$, while the branch of the HP transition with a small temperature only happens in Gauss-Bonnet AdS spacetime with $\alpha_{Tr}<\alpha<\alpha_c$. We also calculate their values in the next subsection.

\subsection{Triple points in $d\geq5$ dimensions}

\begin{figure}[h!]
\begin{center}
\includegraphics[width=0.4\textwidth]{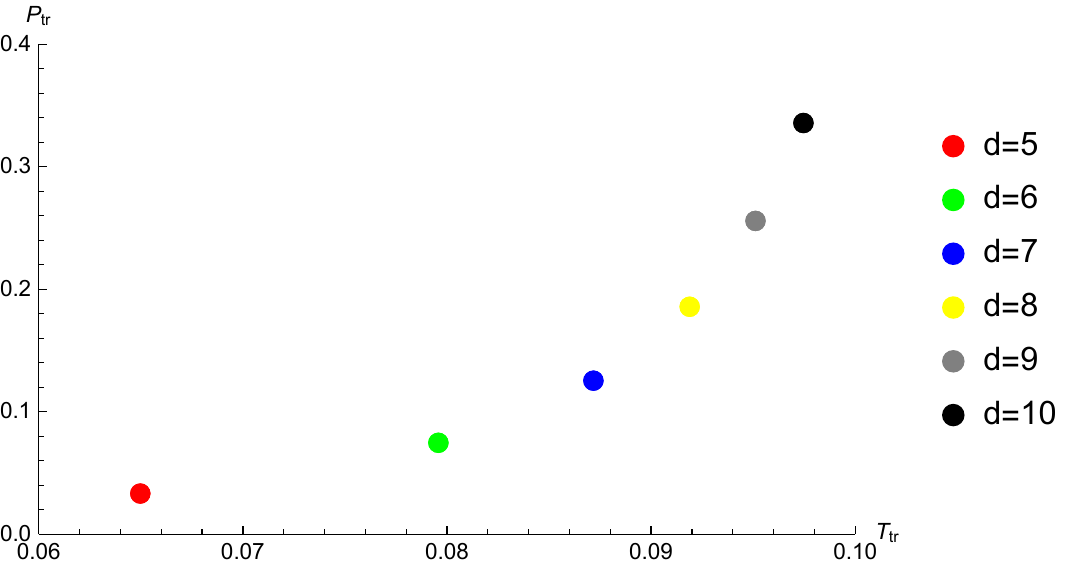}
\includegraphics[width=0.4\textwidth]{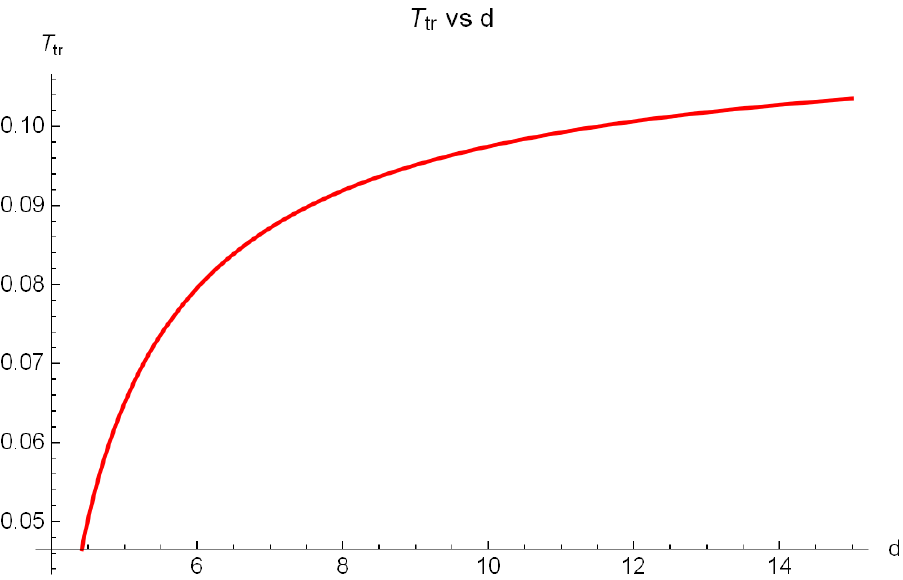}
\caption{Triple points (with $\alpha=1$) in diverse dimensions.}
\label{fig4}
\end{center}
\end{figure}
	
\begin{figure}[h!]
\begin{center}
\includegraphics[width=0.4\textwidth]{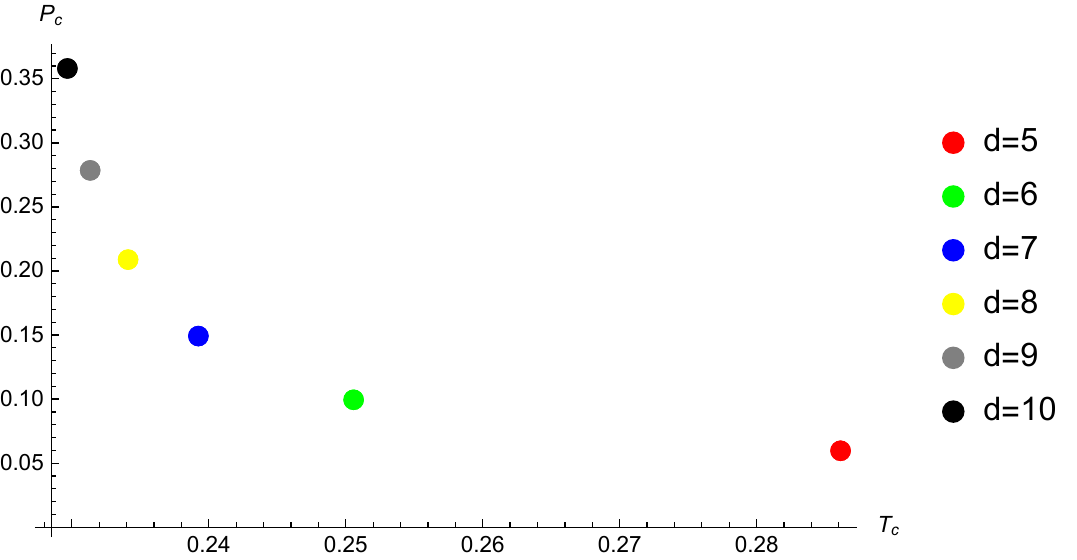}
\includegraphics[width=0.4\textwidth]{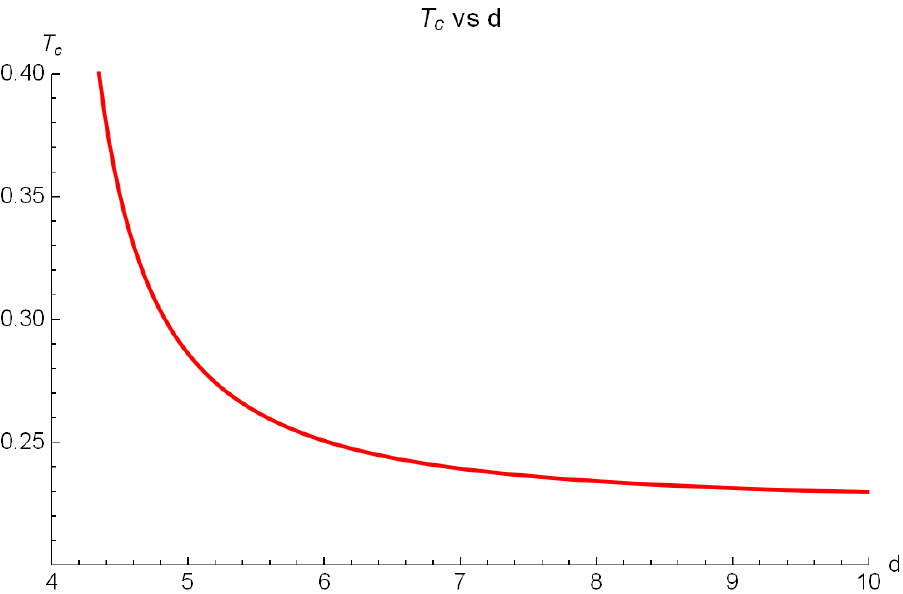}
\caption{Up critical points (with $\alpha=1$) in diverse dimensions.}
\label{fig4-1}
\end{center}
\end{figure}

The triple point (with thermodynamic quantities $(P_{Tr},T_{Tr},r_{Tr})$) corresponds to the black hole phase whose maximum of the Gibbs free energy is zero, which can be calculated by
\begin{align}
  G|_{P=P_{Tr},T=T_{Tr},r=r_{Tr}}=0,\quad\,
  \frac{\partial G}{\partial r_+}\bigg|_{P=P_{Tr},T=T_{Tr},r_+=r_{Tr}}=0.
\end{align}
The former leads to the Eq.(\ref{f0}), while the latter can be simplified as
\begin{equation}
  16 \pi  P R^3 +R^2 \left((d-2)(d-3)-96 \pi  \alpha  P\right) -\alpha  (d-2)(d-9) R +2 \alpha ^2 (d-2)(d-5).
\end{equation}
Combining this two equations and Eq.(\ref{Td}), we can get the triple point
\begin{align}
  P_{Tr}=\frac{d (d-1)(d-4)}{64 (d-2)\pi\alpha},\quad\,T_{Tr}=\frac{1}{2 \sqrt{2} \pi  \sqrt{\frac{\alpha  (d-2)}{d-4}}},\quad\,r_{Tr}=\sqrt{\frac{2 \alpha  (d-2)}{d-4}}.
\end{align}

In order to derive the up critical point with thermodynamic quantities $(P_{c},T_{c},r_{c})$,
we need firstly find the two-phase coexistence state (with ($P_{c},T_{div},r_{div}$)) with a diverging HP temperature, which requires the conditions
\begin{align}
  G|_{P=P_{c},T=T_{div},r_+=r_{div}}=0,\quad\,
  T|_{P=P_{c},r_+=r_{div}}=\infty.
\end{align}
From the latter, one can find directly $r_{div}=\sqrt{2\alpha}$. Inserting it into the former, i.e. Eq.(\ref{f0}),
we obtain the pressure of the up critical point
\begin{align}
  P_{c}=\frac{(d-1)(d-2)}{64\pi\alpha}.
\end{align}
Inserting $P_{c}$ again into Eq.(\ref{f0}) and Eq.(\ref{Td}), we get
\begin{align}
  T_{c}&=\frac{-d^2+\left(\sqrt{d^2-6 d+17}+12\right) d-\sqrt{d^2-6 d+17}-23}{8 \sqrt{2} \pi  \sqrt{\alpha  (d-4) \left(\sqrt{d^2-6 d+17}+3\right)}}\\
  r_{c}&=\sqrt{\frac{2 \alpha  \left( \sqrt{d^2-6 d+17}+3\right)}{d-4}}.
\end{align}

\begin{figure}[h!]
\begin{center}
\includegraphics[width=0.5\textwidth]{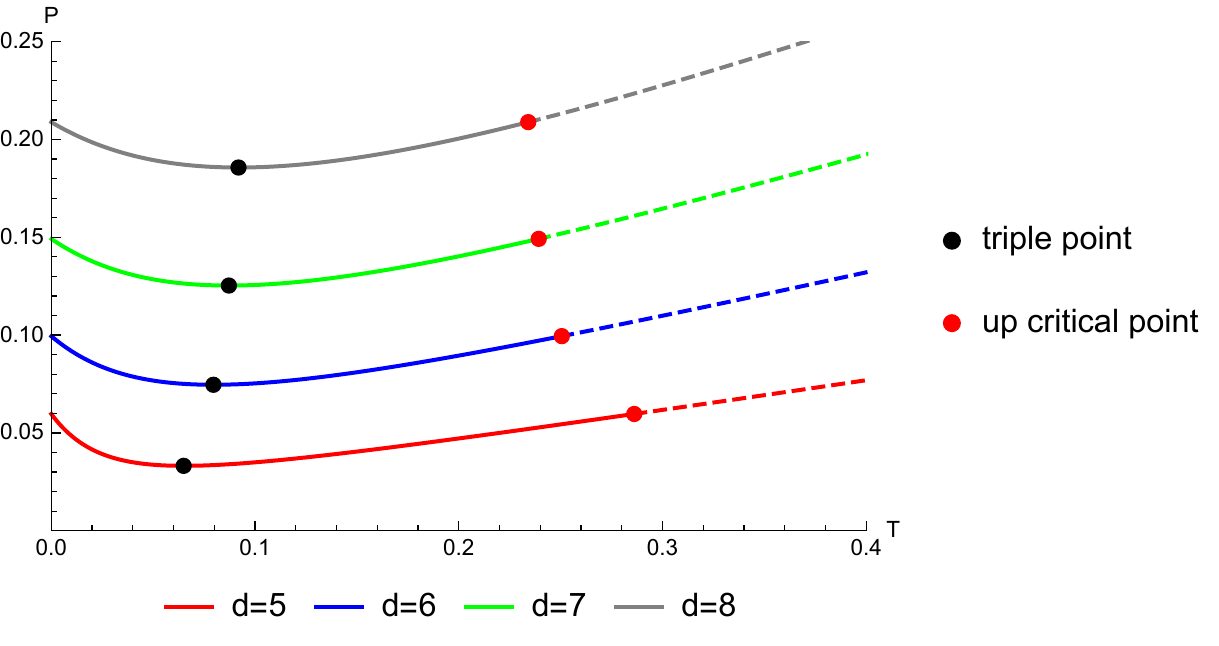}
\caption{$P-T$ phase diagrams (with $\alpha=1$) in diverse dimensions.
The reentrant HP transition are denoted by the solid lines,
while the single HP transition are denoted by the dashed lines.}
\label{fig4-2}
\end{center}
\end{figure}

We show the triple points and up critical points of diverse dimensions in Fig.(\ref{fig4}) and Fig.(\ref{fig4-1}).
When the dimensions $d$ increases, the temperature of the triple point increases while the one of the up critical point
always decreases. We also plot the $P-T$ phase diagrams of diverse dimensions in Fig.(\ref{fig4-2}),
one can always see the Reentrant HP transition behavior of Gauss-Bonnet AdS black holes for
a range of pressure $P\in(P_{Tr},P_c)$ in arbitrary dimensions.
On the other hand, from this range $\frac{d(d-4)(d-1)}{(d-2)64\pi\alpha} <P<\frac{(d-1)(d-2)}{64\pi\alpha}$ for the reentrant HP transition, i.e. $\frac{d(d-4)(d-1)}{(d-2)64\pi} <\alpha P<\frac{(d-1)(d-2)}{64\pi}$,
one can easily conclude that if the pressure $P$ is fixed, the reentrant HP transition will only happen in Gauss-Bonnet AdS spacetime with $\alpha_{Tr}=\frac{d(d-4)(d-1)}{(d-2)64\pi P} <\alpha<\frac{(d-1)(d-2)}{64\pi P}=\alpha_c$.

\subsection{Some universal relations}

\begin{figure}[h!]
\begin{center}
\includegraphics[width=0.5\textwidth]{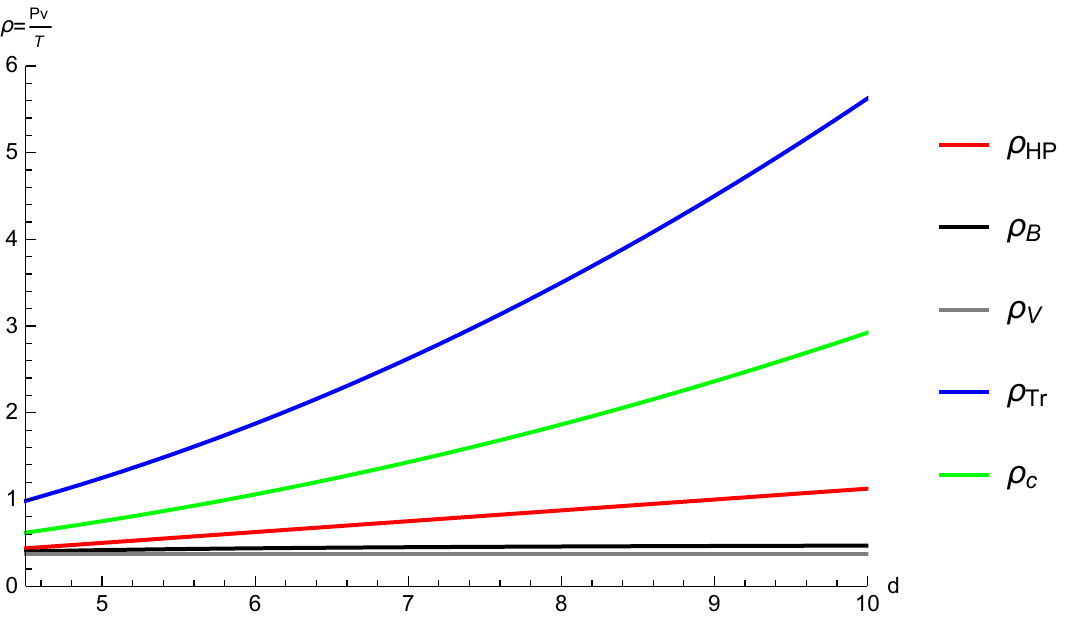}
\caption{Some universal ratios vs dimensions.}
\label{figratio}
\end{center}
\end{figure}

In this subsection, we will demonstrate some universal relations and constants associated with the HP phase transition,
since we expect that they will provide a foundation for understanding black hole thermodynamics,
and other special properties of (other) black holes in AdS spacetime in the quantum and holographic framework.
Actually, universal relations and constants have important applications in understanding a physical theory.

Especially, from the critical phenomenon and phase transition of thermodynamical systems,
some universal relations and constants emerge.
For classical thermodynamical systems, we take the Van der Waals fluid with the equation of state $\bigg(P+\frac{a}{v^2}\bigg)\bigg(v-b\bigg)=kT$ as an example.
From the investigation of the critical point and liquid/vapour phase transition, it is easy to introduce the famous universal relation \begin{align}
  \rho_V=\frac{P_c v_c}{k T_c}=\frac{3}{8}.
\end{align}
As for quantum thermodynamical systems, for instance, black holes,
a similar universal relation
 \begin{align}
  \rho_B=\frac{P_c v_c}{T_c}=\frac{2d-5}{4d-8}
\end{align}
was also found from the investigation of the critical point and SBHs/LBHs phase transition of charged AdS black hole in the extended thermodynamics \cite{Kubiznak:2012wp,Gunasekaran:2012dq}.
These ratios are all universal numbers independent of the parameters of the systems,
e.g. $a,b$ for the Van der Waals fluid and $q$ for charged AdS black hole.

Now we explore the universal relations from the HP transition. Since in the extended thermodynamics of AdS black holes,
the specific volume $v$ is always identified with the horizon radius $r_+$, rather than the thermodynamic volume $V$,
we will study $\rho_{HP}=\frac{P_{HP} r_{HP}}{T_{HP}}$ at the HP transition point with the thermodynamical quantities $(P_{HP}, r_{HP},T_{HP})$. We firstly consider the $d$ dimensional Schwarzschild-AdS black hole.
There exists a HP transition with the thermodynamical quantities $r_{HP}=\sqrt{\frac{(d(d-3)+2)}{16\pi P}}$ and $T_{HP}=\sqrt{\frac{4(d-2)P}{(d-1)\pi}}$. Then we also get a universal relation
\begin{align}
  \rho_{HP}=\frac{P_{HP} r_{HP}}{T_{HP}}=\sqrt{\frac{(d-1)(d(d-3)+2)}{64(d-2)}}.
\end{align}

For the case in our paper, the situation becomes subtle.
We can simplify as $\rho_{HP}=\frac{P_{HP} r_{HP}}{T_{HP}}=\frac{1}{2-\frac{1}{4\pi P r_{HP}^2}}$.
Although $r_{HP}$ has a complicated form, we can insert the four dimensional case Eq.(\ref{rHP}) as an example,
and find an un-universal ratio $\rho_{HP}$ dependent on the Gauss-Bonnet constant $\alpha$.
When we generalize the discussion into the reentrant HP transition, we find the universal ratios for the triple point and up critical point
\begin{align}
  \rho_{Tr}&=\dfrac{P_{Tr}r_{Tr}}{T_{Tr}}=\frac{1}{16} (d-1) d,\\
  \rho_{c}&=\dfrac{P_{c}r_c}{T_{c}}=-\frac{(d-2) (d-1) \left(\sqrt{d^2-6 d+17}+3\right)}{4 \left(d^2-\left(\sqrt{d^2-6 d+17}+12\right) d+\sqrt{d^2-6 d+17}+23\right)}.
  \end{align}
All the above ratios are illustrated in Fig.\ref{figratio}, which are universal and independent of the parameters of the systems.

On the other hand, since the temperature and pressure of the triple points and up critical points
share similar dependence on the Gauss-Bonnet constant $\alpha$, $T\sim \frac{1}{\sqrt{\alpha}}, P\sim \frac{1}{{\alpha}}$, with different $d$-dependent coefficients, we introduce some other universal relations
\begin{equation}
		\begin{aligned}
			\dfrac{T_{Tr}}{T_c}&=-\frac{4 (d-4)(\sqrt{d^2-6 d+17}+3)}{(d^2-\left(\sqrt{d^2-6 d+17}+12\right) (d-1)+23)\sqrt{(d-2)}},\\
			\dfrac{P_{Tr}}{P_c}&=\frac{d(d-4)}{(d-2)^2},
		\end{aligned}
	\end{equation}
which both only depend on the dimensions $d$. 	

\section{Possible relevance and interpretation in the CFT picture}	
\begin{figure}[h!]
\begin{center}
\includegraphics[width=0.75\textwidth]{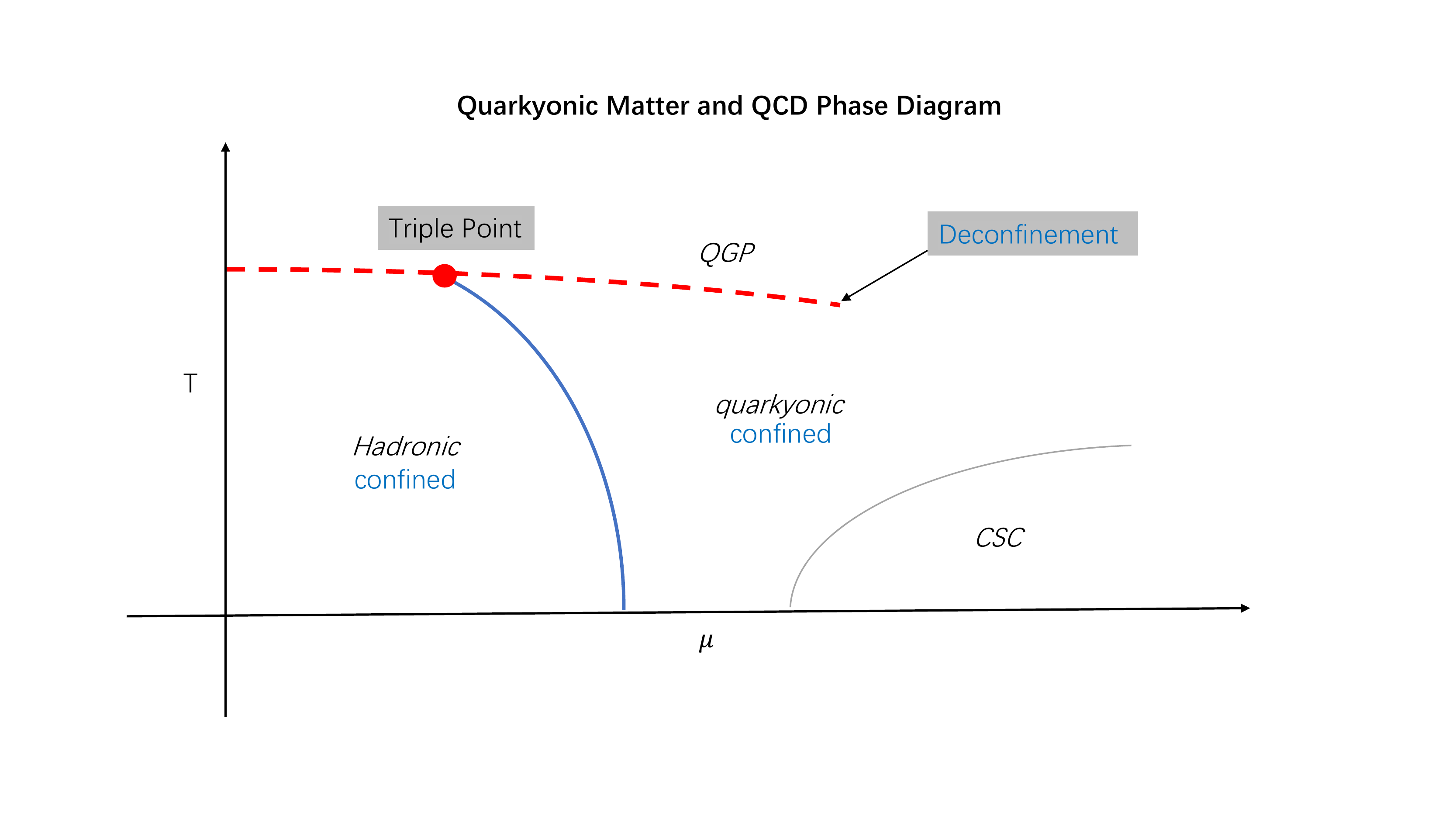}
\caption{Quarkyonic matter, the triple point and the QCD Phase Diagram.}
\label{fig6}
\end{center}
\end{figure}

It is well known that the HP transition could be explained as the confinement/deconfinement phase transition of gauge field \cite{Witten:1998zw}, inspired by the AdS/CFT correspondence \cite{Maldacena:1997re,Gubser:1998bc,Witten:1998qj}.
Thus, do the reentrant HP transition presented in our paper has any possible relevance and interpretation to the AdS/CFT correspondence?
While we can not rigorously prove it, in this section, we conjecture that indeed this is the case.
Here, we conjecture that the reentrant HP transition and the triple point
could be explained as the phase transition and the triple point of gauge field, respectively,
in the QCD phase diagram.

A sketch of a possible phase diagram for QCD is shown in Fig.(\ref{fig6}).
The red dashed line denotes the confirmed confinement/deconfinement phase transition.
Above the transition temperature of the deconfinement phase transition,
QCD is in the Quark-Gluon Plasma phase with deconfinement;
while below the transition temperature, QCD is in the hadronic phase with confinement at small chemical potential $\mu$,
and in the quarkyonic phase with confinement at large $\mu$ \cite{McLerran:2007qj,Hidaka:2008yy,McLerran:2008ua,Fukushima:2008wg,Glozman:2007tv}.
There may exist a quarkyonic phase transition between the hadronic phase and the quarkyonic phase,
below the deconfinement phase transition temperature \cite{McLerran:2007qj}.
The quarkyonic transition is denoted as blue solid line in Fig.(\ref{fig6}).
Lattice QCD simulation also showed the existence of the quarkyonic transition for small $\mu$ \cite{DeTar:2009ef}.
Besides, after considering the Quarkyonic matter,
the two phase transitions and three phases observed in the QCD phase diagram and described above
can be understood as arising from a triple point,
where Hadronic matter, Quarkyonic matter and the Quark-Gluon Plasma all coexist.
The existence of this phase transition and triple point is suggested by many recent studies \cite{Andronic:2009gj,McLerran:2007qj,Doi:2014zea,Pak:2015dxa,Suganuma:2017syi,Yang:2020hun,McInnes:2009zp,McInnes:2009ux,McInnes:2010ti,McInnes:2012bt,Ong:2014maa,McInnes:2015pya,McInnes:2015hga,McInnes:2016dwk,McInnes:2018mwj,Henriksson:2019ifu}.
Noting that at low temperature and very large chemical potential, QCD is conjectured to be in the color superconductor.

The deconfinement phase transition is of first order while the dual HP transition is also a first order phase transition.
The deconfinement phase transition reduces to a crossover \cite{Fromm:2011qi}, even for finite $N_c$ \cite{Li:2018ygx},
which implies a critical endpoint at certain $\mu$.
This crossover is speculated that, in the holographic dictionary,
corresponds to the HP crossover of Schwarzschild AdS black hole \cite{Nicolini:2011dp} in noncommutative spacetime
which is thought to be an effective description of quantum gravitational spacetime.
Similarly, since the special phase transition in QCD phase diagram is composed of the deconfinement phase transition
and the quarkyonic transition, it is spontaneous to speculate that
it may correspond to the reentrant HP transition consisting of two branches of HP transition;
meanwhile the triple point that SBHs/MBHs/LBHs all coexist may be explained as the triple point that Hadronic matter/Quark-Gluon Plasma/Quarkyonic matter all coexist.
On the other hand, the duality between the deconfinement phase transition and HP transition is exactly established  \cite{Witten:1998zw} in Einstein gravity which is just the first term in an infinite series of
gravitational corrections built from powers of the curvature tensor and its derivatives in string theory.
Therefore, up to the second order in the curvature, in Gauss-Bonnet gravity,
the duality between the reentrant HP transition and the Hadronic matter/Quark-Gluon Plasma/Quarkyonic matter transition,
and the duality between these triple points (in black hole thermodynamics and in QCD),
shed some light on black hole thermodynamics and the AdS/CFT correspondence in the quantum gravity picture.

However, giving something more quantitative rigorously here is very difficult,
since string loops can not be yet calculated in any of the backgrounds thought to be dual to gauge theory.
Miserably, the duality between the first order phase transitions is presented on spherically compactified spaces,
while the reentrant HP transition is dealt for Gauss-Bonnet AdS black hole with the horizon being hyperbolic topology.
This fact makes it of importance to explore the reentrant HP transition for the spherical AdS black holes,
which is left as a future task.
The vantage is that many studies about HP transition \cite{Mbarek:2018bau,Xu:2019xif,Aharony:2019vgs,Wu:2020tmz,Wang:2019vgz,Lala:2020lge,Wang:2020pmb,Zhao:2020nrx,Yan:2021uzw,Su:2021jto,Du:2021wvt},
especially in the microcosmic \cite{Li:2020khm,Xu:2020ubo,Li:2021zep}, holographic \cite{Copetti:2020dil,Wei:2020kra,Belhaj:2020mdr} and AdS/QCD \cite{McInnes:2009zp,McInnes:2009ux,McInnes:2010ti,McInnes:2012bt,Ong:2014maa,McInnes:2015pya,McInnes:2015hga,McInnes:2016dwk,McInnes:2018mwj,Henriksson:2019ifu} framework, are aroused, very recently, which are useful for the future investigations.

\section{Conclusion}
In this paper, we investigate HP transition of hyperbolic AdS black hole in extended thermodynamics of $d$ dimensional Gauss-Bonnet gravity. When $d\geq4$, a new family of HP transition, i.e. the reentrant HP transition is found for the first time, which is composed of two HP transitions with a large and a small HP temperature.
We also find the triple point that SBHs/MBHs/LBHs all coexist.
We calculate the temperature of two branches of HP transitions, which both depend on the pressure (i.e. the cosmological constant) and the Gauss-Bonnet constant. It is shown that pressure $P$ and the Gauss-Bonnet constant $\alpha$ both enlarge the large
HP temperature and diminish the small HP temperature.
We present the $P-T$ phase diagrams of Gauss-Bonnet AdS black hole.
We find that the reentrant HP transition always exists in four dimensions, while in $d>4$ dimensions it can only be seen for
a range of pressure $P\in(P_{Tr},P_c)$, which is just the pressure of the triple point and an up critical point.
Above the pressure (temperature) of the up critical points, the $d>4$ dimensional Gauss-Bonnet AdS black hole systems undergo a single HP transition. The triple points and the up critical points for arbitrary dimensional Gauss-Bonnet AdS black hole systems are given, together with some interesting universal relations which only depend on the dimensions $d$ and are independent of the parameters of the systems.

We have speculated the possible relevance and interpretation of the reentrant HP transition and the triple point in the CFT picture.
We conjecture that the reentrant HP transition and the triple point for SBHs/MBHs/LBH could be explained
as the phase transition and the triple point of Hadronic
matter/Quark-Gluon Plasma/Quarkyonic matter in QCD phase diagram.
These results in this paper may improve the comprehension of the black hole thermodynamics in the quantum gravity framework and shed some light on the AdS/CFT correspondence beyond the classical gravity limit.
However, it is really difficult to give some quantitative rigorously evidences.
This fact makes it important to explore more reentrant HP transitions in different backgrounds,
and generalize the study into the microcosmic and holographic framework.
For example, it is interesting to consider the effect of the general higher-derivative terms on the HP transition of
the AdS black holes \cite{Xu:2013zea,Xu:2014tja,Xu:2014kwa,Xu:2018fag,Wu:2021zyl,Akbarieh:2021vhv,Lin:2020kqe,Wei:2020poh,Yang:2020czk} and non-Schwarzschild black holes \cite{Lu:2015cqa,Kehagias:2015ata,Lu:2015psa,Pravda:2016fue,Podolsky:2018pfe,Svarc:2018coe}, as well as the charged and rotating black holes, in AdS spacetime.

\section*{Acknowledgements}
We would like to thank Huan Chen, Cong-yu Wang and Zi-qiang Zhang for useful conversations.
Wei Xu is supported by the National Natural Science Foundation of China (NSFC)
and the Fundamental Research Funds for the Central Universities, China University of Geosciences (Wuhan).
Bin Zhu is supported by the National Science Foundation
of China (NSFC) under Grant No.11747026 and 11805161, and the Natural Science
Foundation of Shandong Province under Grant No.ZR2018QA007.

\providecommand{\href}[2]{#2}\begingroup
\footnotesize\itemsep=0pt
\providecommand{\eprint}[2][]{\href{http://arxiv.org/abs/#2}{arXiv:#2}}

\end{document}